\documentclass[american,twocolumn, showpacs,prl,floatfix]{revtex4}
\usepackage[T1]{fontenc}
\usepackage[latin1]{inputenc}
\usepackage{graphicx}
\usepackage{amssymb}
\usepackage{slashed}
\usepackage{color}
\usepackage{babel}

\newcommand{\vv}[1]{\mathbf{#1}}

\newcommand{\tauiso}{\tau_{\text{iso}}}
\newcommand{\phard}{p_{\text{hard}}}

\newcommand{\todo}[1]{}
\renewcommand{\todo}[1]{{\bf #1}}

\begin{document}

\title{Yoctosecond metrology through HBT correlations from a quark-gluon plasma}

\date{July 1, 2012}

\author{Andreas \surname{Ipp}}
\email{ipp@hep.itp.tuwien.ac.at}

\author{Peter \surname{Somkuti}}
\email{somkuti@hep.itp.tuwien.ac.at}

\affiliation{Institute for Theoretical Physics, Vienna University of Technology, Wiedner Hauptstr. 8-10/136, A-1040 Vienna, Austria}

\begin{abstract}
Expansion dynamics at the yoctosecond timescale affect
the evolution of the quark gluon plasma (QGP)
created in heavy ion collisions.
We show how these
dynamics are accessible through Hanbury Brown and Twiss (HBT) intensity
interferometry of direct photons emitted from the interior of the QGP.
A detector placed close to the beam axis is particularly sensitive to early
polar momentum anisotropies of the QGP.
Observing a modification of
the HBT signal at the proposed FoCal detector of the LHC ALICE experiment
would allow to measure the isotropization time of the plasma
and could provide first experimental evidence for photon double
pulses at the yoctosecond timescale.
\end{abstract}


\maketitle

The quark-gluon plasma (QGP) as created in heavy ion colliders like RHIC or LHC 
exists for a duration of a few tens of yoctoseconds
(1 ys = $10^{-24}\,\mathrm{s}$).
During its expansion, the plasma is exposed to different momentum anisotropies.
Azimuthal anisotropies arise in non-central collisions
and are responsible for the elliptic flow \cite{Adler2003,Aamodt2010}.
Polar anisotropies arise
at very early times right after the collision due to the longitudinal expansion of the plasma
\cite{Blaizot:2011xf}.
The latter kind of anisotropies
causes a variety of fascinating effects:
early polar momentum space anisotropies can induce
Chromo-Weibel plasma instabilities~\cite{Weibel:1959zz,Mrowczynski:1993qm,Romatschke:2003ms,Arnold:2003rq,Ipp:2010uy},
could allow for a violation of the viscosity bound~\cite{Rebhan:2011vd},
or can lead to photon double pulses that are separated merely by yoctoseconds~\cite{Ipp:2009ja}.

Photons are a particularly suitable probe for the early phase of the plasma,
because
once they are produced through quark Compton scattering or quark-antiquark annihilation,
they leave the strongly interacting plasma likely without further interaction.
It turns out that the photon production process is strongly polarization and direction dependent~\cite{Schenke2007,Ipp2008}.
A strong polar anisotropy can lead to temporary suppression
of photon emission in forward direction and thus to non-trivial pulse shapes
that differ from the decay one would expect from an isotropically cooling plasma.
Under particular conditions, even double pulses seem possible~\cite{Ipp:2009ja}.
Measuring the pulse envelope of photons on the yoctosecond timescale would therefore provide firsthand information 
about this early evolution.

Unfortunately, there are no detectors available yet that could time-resolve a possible signal at the yoctosecond scale.
State-of-the-art laser physics deals with attosecond metrology \cite{Hentschel:2001,Krausz2009}.
The next generation of laser facilities like the Extreme Light Infrastructure (ELI)~\cite{ELI}
or the IZEST \cite{IZEST} strive to produce zeptosecond photon pulses.
Even though there are suggestions to characterize photon pulses down to zeptosecond timescale~\cite{Ipp:2010vk},
their applicability to photons from the QGP seems doubtful.
A feasible way to resolve the space-time dynamics at the femtometer and yoctosecond scale
is the Hanbury Brown and Twiss (HBT) interferometry.
Different from intensity interferometry of hadrons (like $\pi$ or $\eta$) which essentially
probe the surface of the plasma, photons provide information from the interior of the plasma~\cite{Neuhauser,Srivastava1993b}.
Previous calculations of photon interferometry for central \cite{Bass2004,Renk2005} and non-central \cite{Frodermann:2009nx}
heavy-ion collisions,
including effects of extremely strong magnetic fields \cite{Itakura:2012dy},
have assumed isotropic photon emission even at the very early stages of the QGP.

\begin{figure}[t]
\centering
\includegraphics[width=7.5cm]{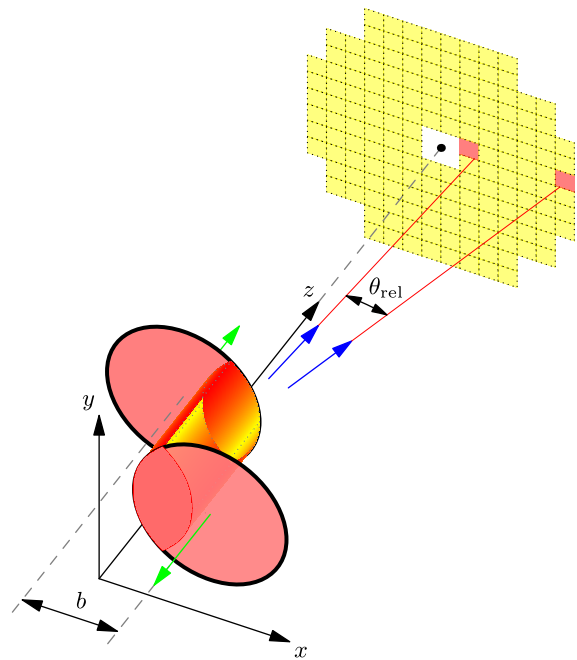}
\caption{\label{fig-system}(color online) 
Concept of HBT detection.
Two heavy ions collide with impact parameter $b$
and produce a QGP that rapidly expands at early times along the beam axis ($z$-axis).
Photons are emitted from the QGP and arrive at a detector placed in forward direction,
like the proposed FoCal detector \cite{Peitzmann:2011ma}.
HBT correlations are obtained from photons that arrive simultaneously,
whose momentum vectors are separated by an angle $\theta_\mathrm{rel}$.
}
\end{figure}

In this Letter, for the first time, we take into account early polar momentum space anisotropies for the photon emission rates to calculate two-photon momentum correlations.
Modifications of the correlation function of photons emitted
close to the beam axis
can be linked to a non-trivial temporal evolution
of the photon emission due to polar anisotropy.
We show that the detection of intensity correlations close to the beam axis,
as illustrated in Fig.~\ref{fig-system}, will be feasible
in the future Forward Calorimeter (FoCal) detector which is likely to be installed in 2017/18
during the ALICE detector upgrade~\cite{Peitzmann:2011ma}.
Our method would allow to measure the isotropization time of a QGP,
and could also establish experimentally the existence of
photon double pulses at the yoctosecond timescale.

The photons of a few GeV energy we want to detect are predominantly produced at very early times,
within a few yoctoseconds after the collision.
In the description of correlation functions, we can thus neglect contributions
from radiative decays of long-lived hadrons like $\pi^0$ and $\eta$, 
because of much larger length and time scales involved which translate to relative momenta
that are smaller by orders of magnitude \cite{Bass2004}.
Also, we neglect a transverse expansion of the system, since the high-energy photons
are most likely emitted close to the center of the QGP.

The HBT correlation function for two photons with momenta $\vv{k}$ and $\vv{k}'$ is given by $C_2(\vv{k},\vv{k}')=P_2(\vv{k},\vv{k}') / P_1(\vv{k})P_1(\vv{k}')$, where $P_1(\vv{k}) = \int d^4x\, w(x,k)$ is the single-particle 
and $P_2(\vv{k},\vv{k}')$ the two-particle inclusive distribution function
\cite{Timmermann1994}
\begin{eqnarray}
\label{eqn:numhbt2}
P_2(\vv{k},\vv{k}') & = & \int d^{4}x\,d^{4}x'\,w\left(x,\frac{k+k'}{2}\right) w\left(x',\frac{k+k'}{2}\right) \nonumber \\
 & & \times \left[ 1+\frac{1}{2}\cos(\Delta k \cdot \Delta x) \right]
\end{eqnarray}
with $\Delta k = k - k'$ and $\Delta x = x-x'$. The factor $\frac{1}{2}$ is a statistical factor from averaging over the photon spin~\cite{Slotta1997}. 
The source function $w(x, k)=dR(x,\vv{k})/(d^{4}x\,d^{3}k)$ describes the mean number of
particles of four-momentum $k$ emitted from a source element centered at the space-time
point $x$~\cite{Timmermann1994}.
A chaotic source is assumed, as contributions by correlated two-photon emissions
are estimated to be negligible \cite{Bass2004}.

In the framework of the current formalism \cite{Neuhauser} one can show
analytically that a source function composed
of two temporally separated Gaussians
leads to oscillations in momentum space in the HBT functions.
However, oscillations in the correlation functions could
also be caused by two spatially separated emission centers.
The question arises how to distinguish a temporal variation from a spatial variation in the HBT signal.
This is possible
by combining correlation measurements in different directions:
HBT oscillations due to temporal separation are independent of the direction of observation,
while spatially separated sources lead to a strong directional dependence of the HBT signal.

In practice, emission centers will not follow a perfect Gaussian shape.
It has been found that the two-photon correlator shows a strongly
non-Gaussian shape along the polar direction \cite{Frodermann:2009nx}.
In the azimuthal direction, oscillations can appear due to a
non-circular intersection region in non-central collisions~\cite{Frodermann:2009nx}.
In the following we present evidence that early polar momentum space anisotropies
can lead to observable modifications of the HBT signal, including the appearance of a side peak,
although we are aware that it will be experimentally challenging
to distinguish such modifications from other possible sources of
oscillations in the HBT correlation functions.
Note that there is also the possibility of introducing 
fake oscillations due to inaccurate numerical integrations~\cite{Timmermann1994,Aichelin:1996iu}.
We therefore carefully cross-checked our numerical results presented below using analytical
and semi-analytical models of temporally separated sources.

The longitudinal expansion of a plasma right after a collision can be described in
two limiting cases:
one is the ideal hydrodynamical evolution as described in the Bjorken expansion picture~\cite{Bjorken1983},
where quark and gluon distribution functions stay isotropic
throughout the expansion.
The other extreme is the free streaming limit~\cite{Kampfer1994}, which neglects all parton interactions, and where momentum anisotropy increases over time.
We base our calculation on a model that can interpolate between these two limiting cases.
The momentum anisotropy is implemented through a modification of isotropic distribution functions $f_{\text{iso}}$ according to 
\cite{Romatschke2003}
\begin{equation}
f(\vv{p}) = f_{\text{iso}}(\sqrt{\vv{p}^2 + \xi(\vv{p}\cdot \hat{\vv{n}})^2}),
\end{equation}
where $\hat{\vv{n}}$ points along the beam axis and the anisotropy parameter
$\xi = \langle p_T^2 \rangle/(2 \langle p_L^2 \rangle)-1$
is defined in the range $-1 < \xi < \infty$.
Values of $\xi > 0$ contract an isotropic distribution along the beam axis, while $\xi < 0$ stretch it.
A single parameter $\delta$ can describe the scaling solutions of ideal hydrodynamical evolution ($\delta=0$) and free-streaming expansion ($\delta=2$), as well as other expansion scenarios like momentum-space broadening due to interactions ($\delta=2/3$)~\cite{Baier2001}.
The time evolution of the anisotropy parameter $\xi$ or the hard momentum scale $\phard$,
which plays the role of temperature in an anisotropic medium, is then given by
\begin{eqnarray}
\xi(\tau) & = &\left( \frac{\tau}{\tau_0} \right)^\delta - 1,\\
\phard(\tau) & = &T_0 \left( \frac{\tau_0}{\tau} \right)^{(1-\delta/2)/3}.
\end{eqnarray}
We follow the model of the plasma evolution of Ref.~\cite{Martinez2008b} who introduced a smeared
step function $\lambda(\tau, \tau_\mathrm{iso}, \gamma)$. By basically replacing $\delta\rightarrow \delta (1-\lambda(\tau, \tau_\mathrm{iso},\gamma))$, this function governs the change of $\delta$ from e.g. 2 or 2/3 to $\delta = 0$
at approximately the isotropization time $\tauiso$ with a parameter $\gamma$
which determines the smoothness of the transition.
The initial temperature distribution for central and non-central collisions
is assumed to be proportional to the thickness functions of the colliding nuclei
according to the Glauber model~\cite{glauber}.
As in Ref.~\cite{Fries2003}, we use hard spheres to model the initial nuclear charge density.

The main contributions of photon production arise from quark-Compton scattering and 
quark-antiquark annihilation processes \cite{Kapusta1991}. 
In an anisotropic plasma, the corresponding photon production rate
$Ed^3R/d^3k$ can only be obtained numerically and shows strong
directional dependence~\cite{Schenke2007,Ipp2008}.
Corresponding expressions for bremsstrahlung
or inelastic pair annihilation become important at lower energies.
Therefore, as in Ref.~\cite{Schenke2007}, we do not take into account these soft scattering processes.
As in Ref.~\cite{Ipp:2009ja}, we use parameters that are relevant to heavy ion collisions at the LHC, with
initial temperature $T_0 = 845\,\text{MeV}$, plasma freezeout temperature $T_c=160\,\text{MeV}$,
nucleus radius $R=7.1\,\text{fm}$, and a plasma formation time of $\tau_0=0.088\,\text{fm}/c$.
To estimate the effect of early polar momentum anisotropies,
various isotropization times starting from ideal hydrodynamic expansion ($\tau_\mathrm{iso}=\tau_0$)
up to ideal free-streaming $\delta = 2$ with isotropization time $\tau_\mathrm{iso}=2\,\mathrm{fm}/c$
are compared.

\begin{figure}
\centering
\includegraphics[width=0.95\columnwidth]{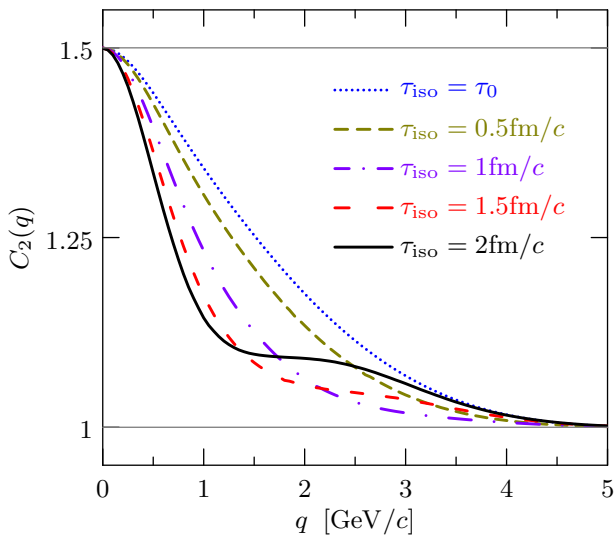}
\caption{\label{fig:collinear}(color online) HBT photon correlation
for a collinear configuration.
The detector is placed at a fixed angle $\theta=25^\circ$ ($\eta\approx1.5$)
and the difference $\mathbf{q} = \mathbf{k}' - \mathbf{k}$
between collinear photon momenta $\mathbf{k}' \parallel \mathbf{k}$
with $k=4\,\mathrm{GeV}/c$ ($k_T \approx 1.7\,\mathrm{GeV}/c$)
is varied.
The various isotropization timescales correspond to
initial free streaming expansion with intermediate polar momentum anisotropy
starting from $\tau_\mathrm{iso}=2\,\mathrm{fm}/c$ (solid line)
down to $\tau_\mathrm{iso}=\tau_0$ (dotted line) which corresponds to
an ideal hydrodynamic expansion.
The impact parameter is chosen as $b=10$ fm and the anisotropy model parameter as $\gamma = 2$.
}
\end{figure}

We consider two different configurations for the calculation of the HBT
correlation function: a collinear and a non-collinear configuration.
In the collinear configuration in Fig.~\ref{fig:collinear}, the two photon momentum
vectors $\mathbf{k}$ and $\mathbf{k}'$ point into the same direction
($\mathbf{k} \parallel \mathbf{k}'$) at a polar angle $\theta$ away from
the beam axis. As in \cite{Srivastava1993b,Timmermann1994},
the free parameter of the two-particle HBT function is the
momentum difference $\mathbf{q} = \mathbf{k}' - \mathbf{k}$.
We focus on collisions with a large impact parameter $b\gtrsim 10$ fm
so that a possible signal is not averaged out by the
transverse size of the plasma~\cite{Ipp:2009ja}.
For hydrodynamic expansion ($\tau_{\mathrm{iso}} = \tau_{0}$), the correlation
function shows the unobtrusive behavior of a monotonically decreasing function.
If one includes the effect of early polar momentum anisotropies
with $\tau_{\mathrm{iso}} \gtrsim 1.5\,\mathrm{fm}/c$ however,
the HBT function reveals a non-trivial shape and
exhibits a plateau-like structure. 
The reason for the appearance of such plateaus is the modification of
the photon production rate in an anisotropic plasma.
The suppression of the photon production in forward direction at
times $\tau_{0} < \tau < \tau_{\mathrm{iso}}$ before isotropization
leads to a non-trivial emission envelope which
could result in two temporally separated peaks \cite{Ipp:2009ja}.
Such an emission envelope leads to a side peak in the correlation function.
A drawback of this configuration is that collinear photons can not be readily
resolved by photon calorimeters.

\begin{figure}
\centering
\includegraphics[width=0.95\columnwidth]{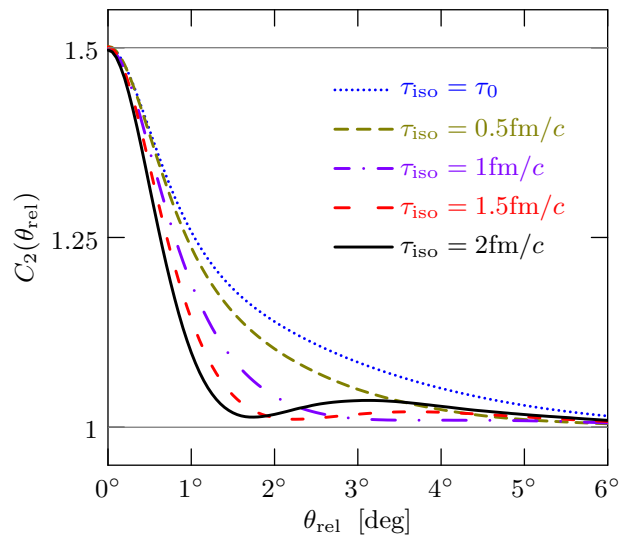}
\caption{\label{fig:noncollinear}(color online) HBT photon correlation for a detector
placed in forward direction.
One momentum vector is placed at $2^\circ$ away from the beam axis ($\eta\approx4$)
and the other varied between $2^\circ$ and $8^\circ$ ($\eta\approx4$ to $\eta\approx2.7$,
at the same azimuthal angle) at fixed momentum $k=25\,\mathrm{GeV}/c$.
A distinct difference and change of shape of the HBT function with respect to the
isotropization time of the quark-gluon-plasma is observed.
The impact parameter is $b=12$ fm and the anisotropy model parameter is $\gamma = 2$.
}
\end{figure}

Therefore we also consider a non-collinear
configuration as shown in Fig.~\ref{fig:noncollinear}.
In this case, both photon momenta share the same magnitude
(|$\mathbf{k}| = |\mathbf{k}'|$), but are positioned at a relative angle
$\theta_{\mathrm{rel}}$ to each other
within the reaction plane.
The selected parameter range is covered by the proposed FoCal detector \cite{Peitzmann:2011ma}.
In this forward detector, photons arrive highly blue-shifted.
A fixed photon momentum $k=25\,\mathrm{GeV}/c$ as observed by the detector corresponds to a transverse momentum
which decreases from $k_T\approx3.5\,\mathrm{GeV}/c$ at $\eta\approx2.7$ to
$k_T\approx0.9\,\mathrm{GeV}/c$ at $\eta\approx4$.
Thus, these photons are most likely emitted from the early QGP.
We see similar behavior in the non-collinear HBT
function as in the collinear configuration.
Early polar momentum space anisotropies result in a narrower correlation function as well as the emergence
of a second maximum.
In principle, two temporally separated photon emission peaks would lead to an
oscillation of the correlation function, but beyond the first two peaks, further maxima
are not identifiable for LHC parameters after integration over the space-time evolution of the QGP.
The distance between the main peak and the side peak of the correlation function
is inversely proportional to the time interval between
two temporally separated peaks in configuration space,
and thus also roughly inversely proportional to the
isotropization time.
An assumed isotropization time of $\tau_{\mathrm{iso}} = 2\,\mathrm{fm}/c$
leads to peaks in the correlation function separated by about $3^\circ$
for the parameters chosen.
Such a structure can in principle be observed in the proposed FoCal detector
which will be able to resolve photons that are separated merely by
a fraction of a degree \cite{Peitzmann:2011ma}.

The photon momentum correlations presented here could be influenced by various effects:
Although transverse expansion can be neglected at
very early times close to the center of a collision,
the question is more delicate
for highly non-central collisions. The space-time evolution
in transverse direction may produce additional modifications of the
photon correlation functions.
Also, at very large rapidity one can not assume a boost-invariant
particle multiplicity. Taking this effect into account
will affect the absolute rate of the observed photons, but it will not
destroy the correlation between them.

Regarding the feasibility of the detection, it will be experimentally challenging,
but not impossible.
If one assumes an annual yield of at least $10^6$ prompt photons
in the energy range $k_T=1\,\mathrm{GeV}/c$ to $4\,\mathrm{GeV}/c$
from heavy ion collisions
that hit the FoCal detector \cite{Hori:2011zza}, one would
observe a few hundred photon pairs within the same time frame.
The modification of the signal presented here is only caused
by direct photons and should therefore be distinguishable from background photons.

To summarize, we have calculated the intensity correlation of photons produced
at an early stage of the quark-gluon plasma, taking into account full polar momentum anisotropy.
Besides other known sources that could lead to oscillations in the HBT signal,
we found that the correlation function is particularly sensitive
to the early time evolution of the plasma.
Large isotropization times lead to distinctive modifications of the HBT correlation function
with a side peak appearing a few degrees separated from the main peak.
The detection of such a structure, for example in the proposed FoCal detector at the
ALICE experiment, would enhance our knowledge about the early evolution
of the QGP, including information about the isotropization process and the isotropization time.
It could also provide first indirect experimental evidence for 
possible photon double pulses at the yoctosecond timescale.

\acknowledgements
We thank Dmitri Peressounko and Yuri Kharlov for helpful discussions regarding LHC detector physics, and Pradip Roy for correspondence.

\bibliography{bibliography}

\begin{thebibliography}{36}
\expandafter\ifx\csname natexlab\endcsname\relax\def\natexlab#1{#1}\fi
\expandafter\ifx\csname bibnamefont\endcsname\relax
  \def\bibnamefont#1{#1}\fi
\expandafter\ifx\csname bibfnamefont\endcsname\relax
  \def\bibfnamefont#1{#1}\fi
\expandafter\ifx\csname citenamefont\endcsname\relax
  \def\citenamefont#1{#1}\fi
\expandafter\ifx\csname url\endcsname\relax
  \def\url#1{\texttt{#1}}\fi
\expandafter\ifx\csname urlprefix\endcsname\relax\def\urlprefix{URL }\fi
\providecommand{\bibinfo}[2]{#2}
\providecommand{\eprint}[2][]{\url{#2}}

\bibitem[{\citenamefont{Adler et~al.}(2003)}]{Adler2003}
\bibinfo{author}{\bibfnamefont{C.}~\bibnamefont{Adler}} \bibnamefont{et~al.}
  (\bibinfo{collaboration}{STAR Collaboration}), \bibinfo{journal}{Phys. Rev.
  Lett.} \textbf{\bibinfo{volume}{90}}, \bibinfo{pages}{032301}
  (\bibinfo{year}{2003}).

\bibitem[{\citenamefont{Aamodt et~al.}(2010)}]{Aamodt2010}
\bibinfo{author}{\bibfnamefont{K.}~\bibnamefont{Aamodt}} \bibnamefont{et~al.}
  (\bibinfo{collaboration}{ALICE Collaboration}), \bibinfo{journal}{Phys. Rev.
  Lett.} \textbf{\bibinfo{volume}{105}}, \bibinfo{pages}{252302}
  (\bibinfo{year}{2010}).

\bibitem[{\citenamefont{Blaizot et~al.}(2012)\citenamefont{Blaizot, Gelis,
  Liao, McLerran, and Venugopalan}}]{Blaizot:2011xf}
\bibinfo{author}{\bibfnamefont{J.-P.} \bibnamefont{Blaizot}},
  \bibinfo{author}{\bibfnamefont{F.}~\bibnamefont{Gelis}},
  \bibinfo{author}{\bibfnamefont{J.-F.} \bibnamefont{Liao}},
  \bibinfo{author}{\bibfnamefont{L.}~\bibnamefont{McLerran}}, \bibnamefont{and}
  \bibinfo{author}{\bibfnamefont{R.}~\bibnamefont{Venugopalan}},
  \bibinfo{journal}{Nucl. Phys.} \textbf{\bibinfo{volume}{A873}},
  \bibinfo{pages}{68} (\bibinfo{year}{2012}), \bibinfo{note}{1107.5296}.

\bibitem[{\citenamefont{Weibel}(1959)}]{Weibel:1959zz}
\bibinfo{author}{\bibfnamefont{E.~S.} \bibnamefont{Weibel}},
  \bibinfo{journal}{Phys. Rev. Lett.} \textbf{\bibinfo{volume}{2}},
  \bibinfo{pages}{83} (\bibinfo{year}{1959}).

\bibitem[{\citenamefont{Mrowczynski}(1993)}]{Mrowczynski:1993qm}
\bibinfo{author}{\bibfnamefont{S.}~\bibnamefont{Mrowczynski}},
  \bibinfo{journal}{Phys. Lett.} \textbf{\bibinfo{volume}{B314}},
  \bibinfo{pages}{118} (\bibinfo{year}{1993}).

\bibitem[{\citenamefont{Romatschke and
  Strickland}(2003{\natexlab{a}})}]{Romatschke:2003ms}
\bibinfo{author}{\bibfnamefont{P.}~\bibnamefont{Romatschke}} \bibnamefont{and}
  \bibinfo{author}{\bibfnamefont{M.}~\bibnamefont{Strickland}},
  \bibinfo{journal}{Phys. Rev.} \textbf{\bibinfo{volume}{D68}},
  \bibinfo{pages}{036004} (\bibinfo{year}{2003}{\natexlab{a}}).

\bibitem[{\citenamefont{Arnold et~al.}(2003)\citenamefont{Arnold, Lenaghan, and
  Moore}}]{Arnold:2003rq}
\bibinfo{author}{\bibfnamefont{P.~B.} \bibnamefont{Arnold}},
  \bibinfo{author}{\bibfnamefont{J.}~\bibnamefont{Lenaghan}}, \bibnamefont{and}
  \bibinfo{author}{\bibfnamefont{G.~D.} \bibnamefont{Moore}},
  \bibinfo{journal}{JHEP} \textbf{\bibinfo{volume}{0308}}, \bibinfo{pages}{002}
  (\bibinfo{year}{2003}).

\bibitem[{\citenamefont{Ipp et~al.}(2011{\natexlab{a}})\citenamefont{Ipp,
  Rebhan, and Strickland}}]{Ipp:2010uy}
\bibinfo{author}{\bibfnamefont{A.}~\bibnamefont{Ipp}},
  \bibinfo{author}{\bibfnamefont{A.}~\bibnamefont{Rebhan}}, \bibnamefont{and}
  \bibinfo{author}{\bibfnamefont{M.}~\bibnamefont{Strickland}},
  \bibinfo{journal}{Phys. Rev.} \textbf{\bibinfo{volume}{D84}},
  \bibinfo{pages}{056003} (\bibinfo{year}{2011}{\natexlab{a}}).

\bibitem[{\citenamefont{Rebhan and Steineder}(2012)}]{Rebhan:2011vd}
\bibinfo{author}{\bibfnamefont{A.}~\bibnamefont{Rebhan}} \bibnamefont{and}
  \bibinfo{author}{\bibfnamefont{D.}~\bibnamefont{Steineder}},
  \bibinfo{journal}{Phys. Rev. Lett.} \textbf{\bibinfo{volume}{108}},
  \bibinfo{pages}{021601} (\bibinfo{year}{2012}).

\bibitem[{\citenamefont{Ipp et~al.}(2009)\citenamefont{Ipp, Keitel, and
  Evers}}]{Ipp:2009ja}
\bibinfo{author}{\bibfnamefont{A.}~\bibnamefont{Ipp}},
  \bibinfo{author}{\bibfnamefont{C.~H.} \bibnamefont{Keitel}},
  \bibnamefont{and} \bibinfo{author}{\bibfnamefont{J.}~\bibnamefont{Evers}},
  \bibinfo{journal}{Phys. Rev. Lett.} \textbf{\bibinfo{volume}{103}},
  \bibinfo{pages}{152301} (\bibinfo{year}{2009}).

\bibitem[{\citenamefont{Schenke and Strickland}(2007)}]{Schenke2007}
\bibinfo{author}{\bibfnamefont{B.}~\bibnamefont{Schenke}} \bibnamefont{and}
  \bibinfo{author}{\bibfnamefont{M.}~\bibnamefont{Strickland}},
  \bibinfo{journal}{Phys. Rev.} \textbf{\bibinfo{volume}{D76}},
  \bibinfo{pages}{025023} (\bibinfo{year}{2007}).

\bibitem[{\citenamefont{Ipp et~al.}(2008)\citenamefont{Ipp, Di~Piazza, Evers,
  and Keitel}}]{Ipp2008}
\bibinfo{author}{\bibfnamefont{A.}~\bibnamefont{Ipp}},
  \bibinfo{author}{\bibfnamefont{A.}~\bibnamefont{Di~Piazza}},
  \bibinfo{author}{\bibfnamefont{J.}~\bibnamefont{Evers}}, \bibnamefont{and}
  \bibinfo{author}{\bibfnamefont{C.~H.} \bibnamefont{Keitel}},
  \bibinfo{journal}{Phys. Lett.} \textbf{\bibinfo{volume}{B666}},
  \bibinfo{pages}{315} (\bibinfo{year}{2008}).

\bibitem[{\citenamefont{Hentschel et~al.}(2001)}]{Hentschel:2001}
\bibinfo{author}{\bibfnamefont{M.}~\bibnamefont{Hentschel}}
  \bibnamefont{et~al.}, \bibinfo{journal}{Nature}
  \textbf{\bibinfo{volume}{414}}, \bibinfo{pages}{509} (\bibinfo{year}{2001}).

\bibitem[{\citenamefont{Krausz and Ivanov}(2009)}]{Krausz2009}
\bibinfo{author}{\bibfnamefont{F.}~\bibnamefont{Krausz}} \bibnamefont{and}
  \bibinfo{author}{\bibfnamefont{M.}~\bibnamefont{Ivanov}},
  \bibinfo{journal}{Rev. Mod. Phys.} \textbf{\bibinfo{volume}{81}},
  \bibinfo{pages}{163} (\bibinfo{year}{2009}).

\bibitem[{\citenamefont{Amiranoff et~al.}()}]{ELI}
\bibinfo{author}{\bibfnamefont{F.}~\bibnamefont{Amiranoff}}
  \bibnamefont{et~al.}, \bibinfo{howpublished}{Proposal for a European Extreme
  Light Infrastructure (ELI);
  http://www.extreme-light-infrastructure.eu/pictures/ELI-scientific-case-id17.pdf}.

\bibitem[{\citenamefont{Mourou and Tajima}()}]{IZEST}
\bibinfo{author}{\bibfnamefont{G.}~\bibnamefont{Mourou}} \bibnamefont{and}
  \bibinfo{author}{\bibfnamefont{T.}~\bibnamefont{Tajima}},
  \bibinfo{howpublished}{IZEST Brochure, http://www.int-zest.com/}.

\bibitem[{\citenamefont{Ipp et~al.}(2011{\natexlab{b}})\citenamefont{Ipp,
  Evers, Keitel, and Hatsagortsyan}}]{Ipp:2010vk}
\bibinfo{author}{\bibfnamefont{A.}~\bibnamefont{Ipp}},
  \bibinfo{author}{\bibfnamefont{J.}~\bibnamefont{Evers}},
  \bibinfo{author}{\bibfnamefont{C.~H.} \bibnamefont{Keitel}},
  \bibnamefont{and} \bibinfo{author}{\bibfnamefont{K.~Z.}
  \bibnamefont{Hatsagortsyan}}, \bibinfo{journal}{Phys. Lett.}
  \textbf{\bibinfo{volume}{B702}}, \bibinfo{pages}{383}
  (\bibinfo{year}{2011}{\natexlab{b}}).

\bibitem[{\citenamefont{Neuhauser}(1986)}]{Neuhauser}
\bibinfo{author}{\bibfnamefont{D.}~\bibnamefont{Neuhauser}},
  \bibinfo{journal}{Phys. Lett.} \textbf{\bibinfo{volume}{B182}},
  \bibinfo{pages}{289 } (\bibinfo{year}{1986}), ISSN \bibinfo{issn}{0370-2693}.

\bibitem[{\citenamefont{Srivastava and Kapusta}(1993)}]{Srivastava1993b}
\bibinfo{author}{\bibfnamefont{D.~K.} \bibnamefont{Srivastava}}
  \bibnamefont{and} \bibinfo{author}{\bibfnamefont{J.~I.}
  \bibnamefont{Kapusta}}, \bibinfo{journal}{Phys. Rev.}
  \textbf{\bibinfo{volume}{C48}}, \bibinfo{pages}{1335} (\bibinfo{year}{1993}).

\bibitem[{\citenamefont{Bass et~al.}(2004)\citenamefont{Bass, M\"uller, and
  Srivastava}}]{Bass2004}
\bibinfo{author}{\bibfnamefont{S.~A.} \bibnamefont{Bass}},
  \bibinfo{author}{\bibfnamefont{B.}~\bibnamefont{M\"uller}}, \bibnamefont{and}
  \bibinfo{author}{\bibfnamefont{D.~K.} \bibnamefont{Srivastava}},
  \bibinfo{journal}{Phys. Rev. Lett.} \textbf{\bibinfo{volume}{93}},
  \bibinfo{pages}{162301} (\bibinfo{year}{2004}).

\bibitem[{\citenamefont{Renk}(2005)}]{Renk2005}
\bibinfo{author}{\bibfnamefont{T.}~\bibnamefont{Renk}}, \bibinfo{journal}{Phys.
  Rev.} \textbf{\bibinfo{volume}{C71}}, \bibinfo{pages}{064905}
  (\bibinfo{year}{2005}).

\bibitem[{\citenamefont{Frodermann and Heinz}(2009)}]{Frodermann:2009nx}
\bibinfo{author}{\bibfnamefont{E.}~\bibnamefont{Frodermann}} \bibnamefont{and}
  \bibinfo{author}{\bibfnamefont{U.}~\bibnamefont{Heinz}},
  \bibinfo{journal}{Phys. Rev.} \textbf{\bibinfo{volume}{C80}},
  \bibinfo{pages}{044903} (\bibinfo{year}{2009}).

\bibitem[{\citenamefont{Itakura and Hattori}(2012)}]{Itakura:2012dy}
\bibinfo{author}{\bibfnamefont{K.}~\bibnamefont{Itakura}} \bibnamefont{and}
  \bibinfo{author}{\bibfnamefont{K.}~\bibnamefont{Hattori}}
  (\bibinfo{year}{2012}), \bibinfo{note}{arXiv:1206.3022}.

\bibitem[{\citenamefont{Peitzmann}(2011)}]{Peitzmann:2011ma}
\bibinfo{author}{\bibfnamefont{T.}~\bibnamefont{Peitzmann}}
  (\bibinfo{collaboration}{ALICE Collaboration}), \bibinfo{journal}{J. Phys.}
  \textbf{\bibinfo{volume}{G38}}, \bibinfo{pages}{124128}
  (\bibinfo{year}{2011}).

\bibitem[{\citenamefont{Timmermann et~al.}(1994)\citenamefont{Timmermann,
  Plumer, Razumov, and Weiner}}]{Timmermann1994}
\bibinfo{author}{\bibfnamefont{A.}~\bibnamefont{Timmermann}},
  \bibinfo{author}{\bibfnamefont{M.}~\bibnamefont{Plumer}},
  \bibinfo{author}{\bibfnamefont{L.}~\bibnamefont{Razumov}}, \bibnamefont{and}
  \bibinfo{author}{\bibfnamefont{R.}~\bibnamefont{Weiner}},
  \bibinfo{journal}{Phys. Rev.} \textbf{\bibinfo{volume}{C50}},
  \bibinfo{pages}{3060} (\bibinfo{year}{1994}).

\bibitem[{\citenamefont{Slotta and Heinz}(1997)}]{Slotta1997}
\bibinfo{author}{\bibfnamefont{C.}~\bibnamefont{Slotta}} \bibnamefont{and}
  \bibinfo{author}{\bibfnamefont{U.}~\bibnamefont{Heinz}},
  \bibinfo{journal}{Phys. Lett.} \textbf{\bibinfo{volume}{B391}},
  \bibinfo{pages}{469 } (\bibinfo{year}{1997}).

\bibitem[{\citenamefont{Aichelin}(1997)}]{Aichelin:1996iu}
\bibinfo{author}{\bibfnamefont{J.}~\bibnamefont{Aichelin}},
  \bibinfo{journal}{Nucl. Phys.} \textbf{\bibinfo{volume}{A617}},
  \bibinfo{pages}{510} (\bibinfo{year}{1997}).

\bibitem[{\citenamefont{Bjorken}(1983)}]{Bjorken1983}
\bibinfo{author}{\bibfnamefont{J.~D.} \bibnamefont{Bjorken}},
  \bibinfo{journal}{Phys. Rev.} \textbf{\bibinfo{volume}{D27}},
  \bibinfo{pages}{140} (\bibinfo{year}{1983}).

\bibitem[{\citenamefont{K\"ampfer and Pavlenko}(1994)}]{Kampfer1994}
\bibinfo{author}{\bibfnamefont{B.}~\bibnamefont{K\"ampfer}} \bibnamefont{and}
  \bibinfo{author}{\bibfnamefont{O.~P.} \bibnamefont{Pavlenko}},
  \bibinfo{journal}{Nucl. Phys.} \textbf{\bibinfo{volume}{A566}},
  \bibinfo{pages}{351c} (\bibinfo{year}{1994}).

\bibitem[{\citenamefont{Romatschke and
  Strickland}(2003{\natexlab{b}})}]{Romatschke2003}
\bibinfo{author}{\bibfnamefont{P.}~\bibnamefont{Romatschke}} \bibnamefont{and}
  \bibinfo{author}{\bibfnamefont{M.}~\bibnamefont{Strickland}},
  \bibinfo{journal}{Phys. Rev.} \textbf{\bibinfo{volume}{D68}},
  \bibinfo{pages}{036004} (\bibinfo{year}{2003}{\natexlab{b}}).

\bibitem[{\citenamefont{Baier et~al.}(2001)\citenamefont{Baier, Mueller,
  Schiff, and Son}}]{Baier2001}
\bibinfo{author}{\bibfnamefont{R.}~\bibnamefont{Baier}},
  \bibinfo{author}{\bibfnamefont{A.~H.} \bibnamefont{Mueller}},
  \bibinfo{author}{\bibfnamefont{D.}~\bibnamefont{Schiff}}, \bibnamefont{and}
  \bibinfo{author}{\bibfnamefont{D.~T.} \bibnamefont{Son}},
  \bibinfo{journal}{Phys. Lett.} \textbf{\bibinfo{volume}{B502}},
  \bibinfo{pages}{51} (\bibinfo{year}{2001}).

\bibitem[{\citenamefont{Martinez and Strickland}(2008)}]{Martinez2008b}
\bibinfo{author}{\bibfnamefont{M.}~\bibnamefont{Martinez}} \bibnamefont{and}
  \bibinfo{author}{\bibfnamefont{M.}~\bibnamefont{Strickland}},
  \bibinfo{journal}{Phys. Rev.} \textbf{\bibinfo{volume}{C78}},
  \bibinfo{pages}{034917} (\bibinfo{year}{2008}).

\bibitem[{\citenamefont{Glauber}(1959)}]{glauber}
\bibinfo{author}{\bibfnamefont{R.~J.} \bibnamefont{Glauber}},
  \emph{\bibinfo{title}{Lectures in Theoretical Physics}}
  (\bibinfo{publisher}{New York: Interscience}, \bibinfo{year}{1959}).

\bibitem[{\citenamefont{Fries et~al.}(2003)\citenamefont{Fries, M\"uller, and
  Srivastava}}]{Fries2003}
\bibinfo{author}{\bibfnamefont{R.~J.} \bibnamefont{Fries}},
  \bibinfo{author}{\bibfnamefont{B.}~\bibnamefont{M\"uller}}, \bibnamefont{and}
  \bibinfo{author}{\bibfnamefont{D.~K.} \bibnamefont{Srivastava}},
  \bibinfo{journal}{Phys. Rev. Lett.} \textbf{\bibinfo{volume}{90}},
  \bibinfo{pages}{132301} (\bibinfo{year}{2003}), \eprint{nucl-th/0208001}.

\bibitem[{\citenamefont{Kapusta et~al.}(1991)\citenamefont{Kapusta, Lichard,
  and Seibert}}]{Kapusta1991}
\bibinfo{author}{\bibfnamefont{J.~I.} \bibnamefont{Kapusta}},
  \bibinfo{author}{\bibfnamefont{P.}~\bibnamefont{Lichard}}, \bibnamefont{and}
  \bibinfo{author}{\bibfnamefont{D.}~\bibnamefont{Seibert}},
  \bibinfo{journal}{Phys. Rev.} \textbf{\bibinfo{volume}{D44}},
  \bibinfo{pages}{2774} (\bibinfo{year}{1991}).

\bibitem[{\citenamefont{Hori et~al.}(2011)\citenamefont{Hori, Hamagaki, and
  Gunji}}]{Hori:2011zza}
\bibinfo{author}{\bibfnamefont{Y.}~\bibnamefont{Hori}},
  \bibinfo{author}{\bibfnamefont{H.}~\bibnamefont{Hamagaki}}, \bibnamefont{and}
  \bibinfo{author}{\bibfnamefont{T.}~\bibnamefont{Gunji}}, \bibinfo{journal}{J.
  Phys. Conf. Ser.} \textbf{\bibinfo{volume}{293}}, \bibinfo{pages}{012029}
  (\bibinfo{year}{2011}).

\end{thebibliography}

\end{document}